\documentclass[showpacs,twocolumn,aps]{revtex4}
\usepackage{graphicx}
\usepackage{dcolumn}
\usepackage{bm}
\usepackage{color}
\usepackage[squaren]{SIunits}
\usepackage{amssymb}
\usepackage{natbib}

\begin{document}
%\preprint{4p0mt}

\newcommand{\ie}{{\it i.e.}}
\newcommand{\eg}{{\it e.g.}}
\newcommand{\etal}{{\it et al.}}

%%%%%%%%%%%%%%%%%%%%%%%%%%%% TITLE

\title{Coexistence of superconductivity and antiferromagnetism in single crystals $A_{0.8}Fe_{2-y}Se_2$ (A= K, Rb, Cs, Tl/K and Tl/Rb):
evidence from magnetization and resistivity}

%%%%%%%%%%%%%%%%%%%%%%%%%%%% AUTHORS
\author{R. H. Liu, X. G. Luo, M. Zhang, A. F. Wang, J. J. Ying, X. F. Wang, Y. J. Yan, Z. J. Xiang, P. Cheng, G. J. Ye, Z. Y. Li and X. H. Chen}
\altaffiliation{E-mail: chenxh@ustc.edu.cn\\ } \affiliation{Hefei
National Laboratory for Physical Science at Microscale and
Department of Physics, University of Science and Technology of
China, Hefei, Anhui 230026, People's Republic of China}

\date{\today}

%%%%%%%%%%%%%%%%%%%%%%%%%%%% ABSTRACT

\begin{abstract}

We measure the resistivity and magnetic susceptibility in the
temperature range from 5 K to 600 K for the single crystals
$A$Fe$_{2-y}$Se$_2$ ($A$ = K$_{0.8}$, Rb$_{0.8}$, Cs$_{0.8}$,
Tl$_{0.5}$K$_{0.3}$ and Tl$_{0.4}$Rb$_{0.4}$). A sharp
superconducting transition is observed in low temperature
resistivity and susceptibility, and susceptibility shows 100\%
Meissner volume fraction for all crystals, while an
antiferromagnetic transition is observed in susceptibility at Neel
temperature ($T_N$) as high as 500 K to 540 K depending on A. It
indicates the coexistence of superconductivity and
antiferromagnetism. A sharp increase in resistivity arises from the
structural transition due to Fe vacancy ordering at the temperature
slightly higher than $T_{\rm N}$. Occurrence of superconductivity in
an antiferromagnetic ordered state with so high $T_{\rm N}$ may
suggest new physics in this type of unconventional superconductors.

\end{abstract}

\pacs{74.70.Xa, 74.25.F, 74.23.Ha}

\maketitle

One of the most amazing issues in the correlated electronic system
is that there are usually the coexistence and competing of several
electronic or magnetic orders. High transition temperature ($T_{\rm
c}$) superconducting cuprates have kept being the central topics in
the condensed matter physics in the past 25 years as a result of the
multi-orders, which induced extremely complicated physics.
Especially, the correlation between superconductivity and
antiferromagnetic or spin-density-wave (SDW) order has puzzled the
scientists for decades and has been thought to be related to origin
of high $T_{\rm c}$ superconductivity in the cuprates. The newly
discovered high-$T_{\rm c}$ superconducting iron-pnictides attracted
the worldwide attention immediately after the discovery of
superconductivity\cite{kamihara,chenxh,ZARen} because the
superconductivity occurs proximity to the magnetically ordered state
or more than that, the coexistence of the superconductivity with
antiferromagnetic order\cite{LiuRH,ChenH,Drew}. Naturally, one takes
the iron-pnictides to compare with cuprates, and believes that they
may have the same origin of high $T_{\rm c}$ superconductivity,
which could be closely related to the antiferromagnetism. However,
no consensus has been reached on this issue so far.

Recently, another newly discovered iron-based superconductors with
$A_xFe_{2-y}Se_2$ (A=K, Rb, Cs, Tl) with $T_{\rm c}$ around 30 K are
reported \cite{xlchen,Mizuguchi,Wang,Ying,Krzton,Fang}.
Antiferromagnetic transition can be clearly observed in
magnetization for non-superconducting Tl- or (Tl,K)-intercalated
compound\cite{YingTL,Fang}. Moun-spin rotation/relaxation ($\mu$SR)
 experiments indicate that superconductivity below $T_c=28$ K
microscopically coexists with a magnetic ordering state with the
transition temperature $T_m=478$ K in
$Cs_{0.8}(FeSe_{0.98})_2$\cite{sher}. Very recently, Bao et al.
reported an antiferomagnetism with Neel temperature ($T_N$) as high
as 559 K with the iron magnetic moment of 3.31$\mu_B$, and a
structural transition at $T_s$=578 K due to iron vacancy oedering in
superconducting $K_{0.8}Fe_{1.6}Se_2$\cite{BaoWW}. Iron vacancy
superstructure at $T_s$=500 K and possible antiferromagnetic
ordering with the Fe magnetic moment of 2$\mu_B$ is also reported in
$Cs_xFe_{2-y}Se_2$ (y=0.29 and x=0.83)\cite{Vyu}. It is well known
that there exists a response in resistivity to the magnetic
transition, and the magnetic transition can be detected by the
susceptibility in iron pnictides supercodncutors\cite{WangXF}. In
order to directly study the magnetic transition and to elucidate the
connection between the superconductivity and magnetic order, we
study the high-temperature magnetic susceptibility and resistivity
in the temperature range from 5 K to 600 K, and find the coexistence
of the superconductivity and antiferromagnetism. In this letter, we
report the magnetic susceptibility and resistivity from 5 K to 600 K
for the $A$Fe$_{2-y}$Se$_2$ ($A$ = K$_{0.8}$, Rb$_{0.8}$,
Cs$_{0.8}$, Tl$_{0.5}$K$_{0.3}$ and Tl$_{0.4}$Rb$_{0.4}$). The
antiferromagnetic transition was observed at $T_{\rm N}$ of $\sim$
500-540 K in the magnetic susceptibility for all the superconducting
crystals with 100\% Meissner volume fraction, indicative of the
coexistence of the antiferromagnetism and superconductivity in the
intercalated Iron selenides. A sharp increase in resistivity starts
at the temperature slightly higher $T_{\rm N}$. Such increase of
resistivity could arises from the Fe vacancy ordering.

\begin{figure}[t]
\includegraphics[width = 0.45\textwidth]{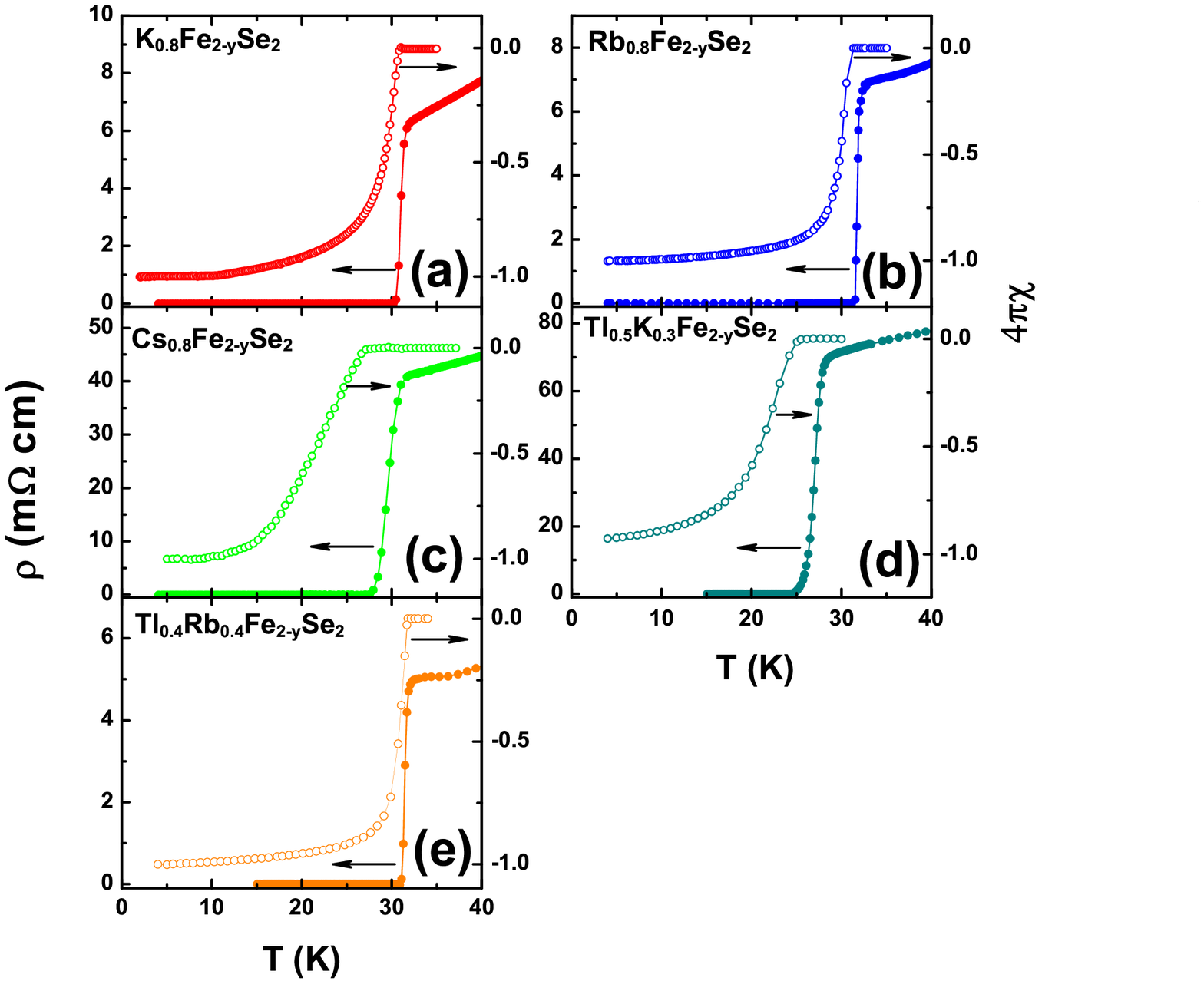}
\caption{(Color online) Temperature dependence of the resistivity
and zero-field-cooled (ZFC) magnetic susceptibility at 10 Oe with
the field applied within {\sl ab}-plane for the superconducting
$A$Fe$_{2-y}$Se$_2$ ($A$ = K$_{0.8}$, Rb$_{0.8}$, Cs$_{0.8}$,
Tl$_{0.5}$K$_{0.3}$ and Tl$_{0.4}$Rb$_{0.4}$) single crystals.}
\end{figure}

The single crystals were grown by Bridgeman method as reported
previously\cite{Ying,Wang}. Resistivity below 400 K was measured
using the {\sl Quantum Design} PPMS-9. The resistivity measurement
above 400 K were carried out with an alternative current resistance
bridge (LR700P) by using the a Type-K Chromel-Alumel thermocouples
as thermometer in a home-built vacuum resistance oven. Magnetic
susceptibility was measured using the {\sl Quantum Design}
SQUID-MPMS. A high-temperature oven was used in the SQUID-MPMS for
magnetic susceptibility measurement above 400 K.

Five systems of superconducting $A$Fe$_{2-y}$Se$_2$ crystals ($A$ =
K$_{0.8}$, Rb$_{0.8}$, Cs$_{0.8}$, Tl$_{0.5}$K$_{0.3}$ and
Tl$_{0.4}$Rb$_{0.4}$) were investigated in this study. The
superconducting transition temperatures ($T_{\rm c}$) for all the
superconducting samples are listed in Table I. As shown in Fig.1,
the superconducting transition width lies between 0.5 to 3 K.
Especially, the transition width for K$_{0.8}$Fe$_{2-y}$Se$_{2}$,
Rb$_{0.8}$Fe$_{2-y}$Se$_{2}$ and
Tl$_{0.4}$Rb$_{0.4}$Fe$_{2-y}$Se$_{2}$ is less than 1 K. The
susceptibility measured in zero-field cooled process at the magnetic
field of 10 Oe shows fully shielding at 5 K for the crystals with
$A$ = K$_{0.8}$, Rb$_{0.8}$, Cs$_{0.8}$, and Tl$_{0.4}$Rb$_{0.4}$,
and 90\% shielding fraction for the crystal with $A$ =
Tl$_{0.5}$K$_{0.3}$.

\begin{figure}[t]
\includegraphics[width = 0.45\textwidth]{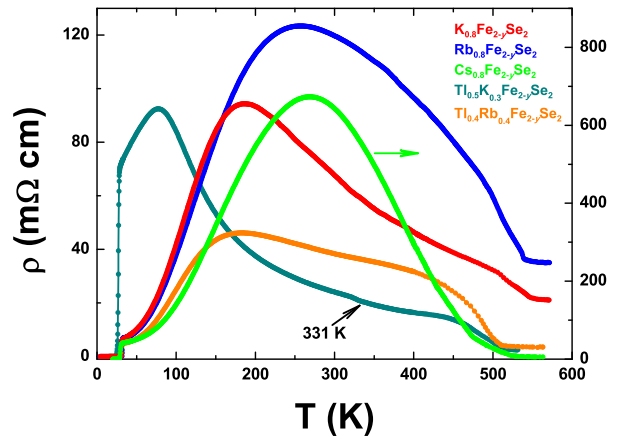}
\caption{(Color online) Temperature dependence of the resistivity
for single crystals $A$Fe$_{2-y}$Se$_2$ )$A$ = K$_{0.8}$,
Rb$_{0.8}$, Cs$_{0.8}$, Tl$_{0.5}$K$_{0.3}$ and
Tl$_{0.4}$Rb$_{0.4}$). The black arrow indicates the kink in the
resistivity.}
\end{figure}

Figure 2 shows the temperature dependence of the resistivity in
temperatures range from 5 K to 600 K for the single crystal
$A$Fe$_{2-y}$Se$_2$ with $A$ = K$_{0.8}$, Rb$_{0.8}$, Cs$_{0.8}$,
Tl$_{0.5}$K$_{0.3}$, and Tl$_{0.4}$Rb$_{0.4}$. All of the samples
display the common features. All samples show superconducting at
$T_{\rm c}$ of $\sim 30$ K, and the $T_c$ is listed in Table I.
Resistivity shows a broad hump in the temperatures range from 70 K
to 300 K ($T_{\rm hump}$) for all crystals. The magnitude of the
resistivity is so high for all the samples compared to the
iron-pnictide superconductors \cite{XFWang,WangXF,HaihWen1} and the
FeSe single crystals\cite{Braithwaite}. Above $T_{\rm hump}$, the
resistivity shows a semiconductor-like behavior. A sharp increase in
resistivity can be observed  above 500 K for all samples, indicative
of the existence of the phase transition.  The temperature ($T_S$),
at which the resistivity starts to sharply increase, varies from 512
to 551 K with changing \emph{A}. Above the $T_S$, the resistivity
shows a weak temperature dependence. The $T_S$ is listed in Table I
for all the samples.

In order to detect the magnetic transition and make clear what
transition inferred by the kinks in the resistivity, we measured the
magnetic susceptibility at 5 T in the temperature range up to 600 K,
as shown in Fig. 3. A pronounced drop is observed in the magnetic
susceptibility at a temperature above 500 K for all the samples. It
indicates the antiferromagnetic transition at these temperatures
($T_{\rm N}$). $T_{\rm N}$ is 540, 534, 504, 500, and 496 K for the
crystals with $A$ =  K$_{0.8}$, Rb$_{0.8}$, Cs$_{0.8}$,
Tl$_{0.5}$K$_{0.3}$ and Tl$_{0.4}$Rb$_{0.4}$, respectively. The
antiferromagnetic order has been found in the
$Cs_{0.8}(FeSe_{0.98})_2$ by $\mu$SR with $T_{\rm N}~\approx$ 478.5
K\cite{sher}. And an antiferromagnetic transition has also been
observed in K$_{0.8}$Fe$_{1.6}$Se$_2$ at $T_N$ as high as 559 K from
Neutron diffraction experiments\cite{BaoWW}. Here, magnetic
susceptibility data indicate the existence of the antiferromagnetic
transition above 490 K for all the crystals with $A$ =  K$_{0.8}$,
Rb$_{0.8}$, Cs$_{0.8}$, Tl$_{0.5}$K$_{0.3}$ and
Tl$_{0.4}$Rb$_{0.4}$. It is worth of noting that the temperature of
the kink in resistivity is slightly higher than those $T_{\rm N}$
observed in the magnetic susceptibility. Actually, $T_{\rm N}$
locates at the middle of the transition observed in the resistivity,
It suggests that the sharp increase in the resistivity at high
temperature is not corresponding to the antiferromagnetic
transition. Indeed, the neutron diffraction results indicate that a
structural transition takes place at a temperature ($T_{\rm S}$)
just above the $T_{\rm N}$ due to the ordering of the iron vacancy,
$T_{\rm N}$ and $T_{\rm S}$ are 559 K and 578 K for the sample
$K_{0.8}Fe_{1.6}Se_2$, respectively\cite{BaoWW}.  It is easily found
that the $T_S$ corresponding to the beginning of the sharp increase
in resistivity is 10-20 K higher than the $T_N$ determined by
susceptibility. Based on the observation by neutron
scattering\cite{BaoWW}, we can infer that the resistivity starts to
sharply increase due to the structural transition, and the kink
temperature can be defined as the structural transition temperature.
Therefore, we can observe the structural and antiferromagnetic
transition from the resistivity and magnetic susceptibility,
respectively.  It should be pointed out that the $T_N$ observed here
in A=K and Cs is different from that reported by Shermadini et
al.\cite{sher} and by Bao et al.\cite{BaoWW}. It could be from
different doping level although their $T_C$ does not change so much.
The black arrow in Fig.3 points out a transition at 332 K in
magnetic susceptibility for the crystal
$Tl_{0.5}K_{0.3}Fe_{2-y}Se_2$. An anomaly can also be observed in
resistivity at 331 K in Fig.2. In fact, a small transition at about
250 K is also observed in susceptibility for the crystal
$Tl_{0.4}Rb_{0.4}Fe_{2-y}Se_2$. Such behavior cannot be observed in
the crystals $A_{0.8}Fe_{2-y}Se_2$ (A=K, Rb, Cs). Such tiny
transition may be due to small amount of Tl$_x$Fe$_{2-y}$Se$_2$
because similar transition has been observed in
Tl$_x$Fe$_{2-y}$Se$_2$ with different $T_{\rm N}$\cite{YingTL,Fang}.

\begin{table*}[t]
\tabcolsep 0pt \caption{Superconducting  transition temperature
($T_{\rm c}^{\rm zero}$, $T_{\rm c}^{\rm onset}$), the hump
temperature in resistivity ($T_{\rm hump}$), antiferromagnectic
transition temperature ($T_{\rm N}$) and structural transition
temperature ($T_{\rm S}$) for the crystals $A$Fe$_{2-y}$Se$_2$ ($A$
= K$_{0.8}$, Rb$_{0.8}$, Cs$_{0.8}$, Tl$_{0.4}$K$_{0.3}$ and
Tl$_{0.4}$Rb$_{0.4}$).} \vspace*{-12pt}
\begin{center}
\def\temptablewidth{\textwidth}
{\rule{\temptablewidth}{1pt}}
\begin{tabular*}{\temptablewidth}{@{\extracolsep{\fill}}cccccc}\hline
 sample name & $T_{\rm c}^{\rm zero}$(K)  & $T_{\rm c}^{\rm onset}$(K) & $T_{\rm hump}$(K) & $T_{\rm N}$(K) & $T_{\rm S}$(K)\\ \hline
      K$$Fe$_{2-y}$Se$_2$ & --- & --- & --- & 527 & 540  \\
      K$_{0.8}$Fe$_{2-y}$Se$_2$ & 30.5 & 31.5 & 170 & 540 & 551  \\
      Rb$_{0.8}$Fe$_{2-y}$Se$_2$ & 31.5 & 32.0 & 250 & 534 & 540 \\
      Cs$_{0.8}$Fe$_{2-y}$Se$_2$ & 27.5 & 30.9 & 270 & 504  & 525 \\
      Tl$_{0.4}$K$_{0.3}$Fe$_{2-y}$Se$_2$ & 24.8 & 27.7 & 78 & 496 &515 \\
      Tl$_{0.4}$Rb$_{0.4}$Fe$_{2-y}$Se$_2$ & 30.9 & 31.8 & 180 & 500 & 512 \\      \hline
       \end{tabular*}
       {\rule{\temptablewidth}{1pt}}
\end{center}
\end{table*}

\begin{figure}[t]
\includegraphics[width = 0.45\textwidth]{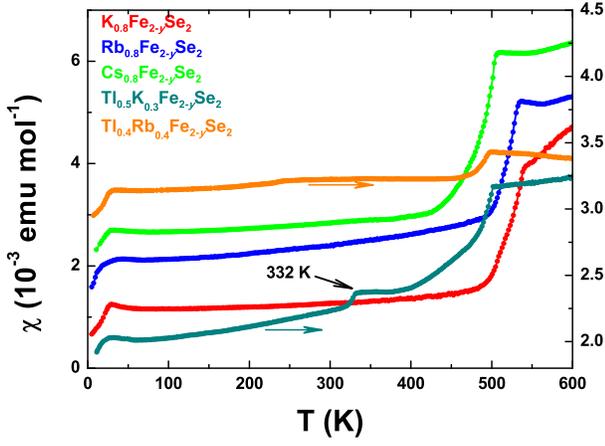}
\caption{(Color online) Magnetic susceptibility measured at 5 T as a
function of temperature for the crystals $A$Fe$_{2-y}$Se$_2$ ($A$ =
K$_{0.8}$, Rb$_{0.8}$, Cs$_{0.8}$, Tl$_{0.5}$K$_{0.3}$ and
Tl$_{0.4}$Rb$_{0.4}$). The black arrow points out a kink in the
susceptibility, and a similar kink is observed in resistivity.}
\end{figure}

\begin{figure}[t]
\includegraphics[width = 0.45\textwidth]{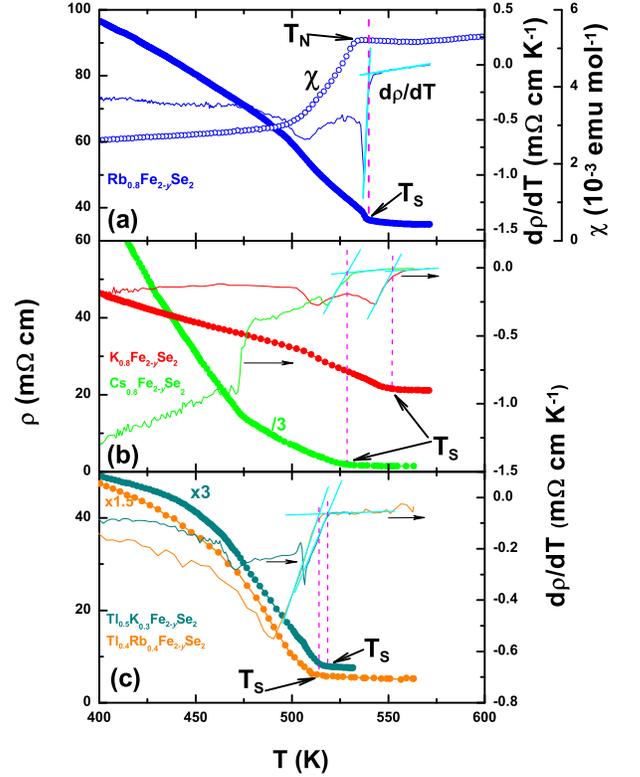}
\caption{(Color online) (a): Comparison of the high-temperature
resistivity, its derivative
 and magnetic susceptibility for crystal Rb$_{0.8}$Fe$_{2-y}$Se$_2$. $T_{\rm s}$
 inferred from the resistivity data and $T_{\rm N}$ inferred from magnetic susceptibility are
 shown. (b) and (c): the high-temperature resistivity and its derivative for single
 crystals $A$Fe$_{2-y}$Se$_2$: (b): $A$ = K$_{0.8}$ and Cs$_{0.8}$; (c): Tl$_{0.5}$K$_{0.3}$ and Tl$_{0.4}$Rb$_{0.4}$.
 $T_{\rm s}$ inferred from the resistivity is shown. }
\end{figure}

In order to carefully determine $T_S$,  the resistivity and the
corresponding derivative (d$\rho$/d$T$) as well as the comparing
with the magnetic susceptibility is plotted in Fig.4a from 400 K to
600 K for the crystal Rb$_{0.8}$Fe$_{2-y}$Se$_2$. A clear kink in
resistivity is observed at 540 K. d$\rho$/d$T$ shows two dips. One
can easily find that the beginning of the high-$T$ dip in
d$\rho$/d$T$ corresponds to the kink in resistivity. This
temperature is defined as $T_{\rm S}$. $T_{\rm N}$ inferred from the
susceptibility is 6 K less than $T_{\rm S}$. It indicates that
$T_{\rm S}$ manifests another phase transition instead of the
antiferromagnetic transition observed in susceptibility. This
transition should be the structural transition due to the ordering
of Fe vacancies because it has been found that the structural
transition occurs just before the magnetic transition\cite{BaoWW}.
$T_{\rm S}$ is determined in the same way for $A$Fe$_{2-y}$Se$_2$
with $A$ = K$_{0.8}$, Cs$_{0.8}$, Tl$_{0.5}$K$_{0.3}$ and
Tl$_{0.4}$Rb$_{0.4}$, as shown Fig. 4b and Fig.4c. The obtained
$T_{\rm S}$ is also listed in Table I. One can find that all $T_{\rm
S}$ is slightly higher than $T_{\rm N}$ in Table I, indicating the
higher transition temperature for the ordering of Fe vacancies than
that of magnetic transition. Therefore, the rapid increase of the
resistivity should be ascribed to arise from the Fe vacancy
ordering, and consequently the very large resistivity in the normal
state in the intercalated iron-selenides originates from the
existence of large amount of Fe vacancies and their ordering. One
can note that the d$\rho$/d$T$ shows two dips for all the
$A$Fe$_{2-y}$Se$_{2}$ crystals except for the
Tl$_{0.4}$Rb$_{0.4}$Fe$_{2-y}$Se$_{2}$. The dip actually manifests
the change of resistivity. Therefore, the second dip can be related
to the occurrence of the antiferromagnetism.

One puzzle in the intercalated iron-selenide single crystals is how
to enter into superconducting state from an antiferromagnetic state
with ordered Fe magnetic moment of 3.3$\mu_B$\cite{BaoWW} and from
the high-temperature semiconductor-like behavior with very high
magnitude of resistivity. One may note that resistivity increases
rapidly below the structural transition temperature. It suggests
that the ordering of the Fe vacancy is responsible for the
semiconductor-like behavior and large magnitude of resistivity above
$T_{\rm hump}$. The resistivity rapidly increases from 21.5 $m\Omega
cm$ to 94.3 $m\Omega cm$ with decreasing temperature from $T_S$ to
$T_{\rm hump}$ for the crystal $A_{0.8}$Fe$_{2-y}$Se$_2$. It
indicates that the Fe vacancy ordering makes the carrier localized
and strongly scatters the charges. Usually, the antiferromagnetic
spin-density-wave transition in the iron-pnictides has been thought
to be related to the reconstruction of Fermi surface (RFS). Such RFS
can induce a more metallic resistivity (like in BaFe$_2$As$_2$ and
LnOFeAs systems). One possible origin of the metallic resistivity
below $T_{\rm hump}$ can be the joint result of the ordering of the
Fe vacancies and the occurrence of antiferromagnetism. All of these
mysteries require further experimental and theoretical study.

The above results indicate that the superconductivity in the
$A_x$Fe$_{2-y}$Se$_2$ happens in an antiferromagnetic ordering state
with very high transition temperature $T_{\rm N}$. In order to
understand the coexistence of the superconductivity and
antiferromagnetic order with very high $T_{\rm N}$, we measured
resistivity and susceptibility on non-superconducting crystal
$KFe_{2-y}Se_2$ (The data are not shown here). Although this sample
is not superconducting and shows a semiconducting behavior in the
whole temperature range, resistivity and magnetic susceptibility
display the similar behavior at high temperatures as those of the
superconducting samples. An antiferromagnetic transition with
$T_{\rm N}$ = 527 K is observed in susceptibility. Surprisingly, the
$T_{\rm N}$ is lower than that in the superconducting
K$_{0.8}$Fe$_{2-y}$Se$_2$ crystal. Resistivity shows a transition at
540 K due to iron vacancy ordering. Both $T_{\rm N}$ and $T_{\rm S}$
are higher in the superconducting sample K$_{0.8}$Fe$_{2-y}$Se$_{2}$
than the non-superconducting sample. It implies that the coexistence
of the antiferromagnetic ordering and the superconductivity is not
simply competing. It requires to clarify how superconductivity
occurs in such antiferromagnetic ordered state.

In summary, we first report the magnetic susceptibility and
resistivity from 5 K to 600 K for the crystals $A$Fe$_{2-y}$Se$_2$
($A$ = K$_{0.8}$, Rb$_{0.8}$, Cs$_{0.8}$, Tl$_{0.5}$K$_{0.3}$ and
Tl$_{0.4}$Rb$_{0.4}$). The structural and antiferromagnetic
transition temperatures are systematically determined by resistivity
and susceptibility for all the superconducting crystals with 100\%
Meissner volume fraction, indicative of the coexistence of the
antiferromagnetism and superconductivity in the intercalated iron
selenides. A sharp increase in resistivity starts at the $T_S$b
slightly higher than $T_{\rm N}$. Such increase of resistivity could
arises from the Fe vacancy ordering.  The higher $T_{\rm N}$ and
$T_{\rm S}$ in the superconducting crystal relative to
non-superconducting crystal suggests that antiferromagnetic
magnetism and superconductiivity are not simply competing to each
other. Occurrence of superconductivity in an antiferromagnetic
ordered state with so high $T_{\rm N}$ and the large magnetic moment
of Fe up to $3.3\mu_B$ may
suggest new physics in this type of unconventional superconductor.\\

{\bf ACKNOWLEDGEMENT:} X. H. Chen would like to thank W. Bao for
uesful discussion. This work is supported by the Natural Science
Foundation of China and by the Ministry of Science and Technology of China, and by Chinese Academy of Sciences.\\

\end{document}